\documentclass[aps,prl, twocolumn, superscriptaddress, floatfix]{revtex4}
\usepackage{amssymb}
\usepackage{amsmath}
\usepackage{graphicx}
\usepackage{color}

\newcommand{\VLP}{V_{\rm LP}}
\newcommand{\VRP}{V_{\rm RP}}
\newcommand{\VL}{V_{\rm L}}
\newcommand{\VR}{V_{\rm R}}
\newcommand{\VM}{V_{\rm M}}

\begin{document}
\title{Double quantum dot with integrated charge sensor based on\\ Ge/Si heterostructure nanowires}
\author{Yongjie~Hu$^{\dagger}$\footnotetext{$^{\dagger}$These authors contributed equally to this work.}
}
\affiliation{Department of Chemistry and Chemical Biology, Harvard University, Cambridge, Massachusetts 02138, USA}
\author{Hugh~O.~H.~Churchill$^{\dagger}$}
\affiliation{Department of Physics, Harvard University, Cambridge, Massachusetts 02138, USA}
\author{David~J.~Reilly}
\affiliation{Department of Physics, Harvard University, Cambridge, Massachusetts 02138, USA}
\author{Jie~Xiang}
\affiliation{Department of Chemistry and Chemical Biology, Harvard University, Cambridge, Massachusetts 02138, USA}
\author{Charles~M.~Lieber{$^{*}$}}
\affiliation{Department of Chemistry and Chemical Biology, Harvard University, Cambridge, Massachusetts 02138, USA}
\affiliation{Division of Engineering and Applied Sciences, Harvard University, Cambridge, Massachusetts 02138, USA}
\email{cml@cmliris.harvard.edu, marcus@harvard.edu}
\author{Charles~M.~Marcus{$^{*}$}}
\affiliation{Department of Physics, Harvard University, Cambridge, Massachusetts 02138, USA}
\date{\today}

%\begin{abstract}
%\end{abstract}

\maketitle

\textbf {Coupled electron spins in semiconductor double quantum dots hold promise as the basis for solid-state qubits \cite{Loss-PRB98,Hanson-condmat06}.  
To date, most experiments have used III-V materials, in which coherence is limited by hyperfine interactions \cite{Petta-Sci05,Koppens-Nat06,Burkard-PRB99,Khaetskii-PRL02}.   
Ge/Si heterostructure nanowires seem ideally suited to overcome this limitation:  the predominance of spin-zero nuclei suppresses the hyperfine interaction and chemical synthesis creates a clean and defect-free system with highly controllable properties \cite{Lu-PNAS05}.  
Here we present a top gate-defined double quantum dot based on Ge/Si heterostructure nanowires with fully tunable coupling between the dots and to the leads. 
We also demonstrate a novel approach to charge sensing in a one-dimensional nanostructure by capacitively coupling the double dot to a single dot on an adjacent nanowire. 
The double quantum dot and integrated charge sensor serve as an essential building block required to form a solid-state spin qubit free of nuclear spin.}

%-------------------%
\begin{figure}[h!]
\center \label{figure1}
\includegraphics[width=3.4in]{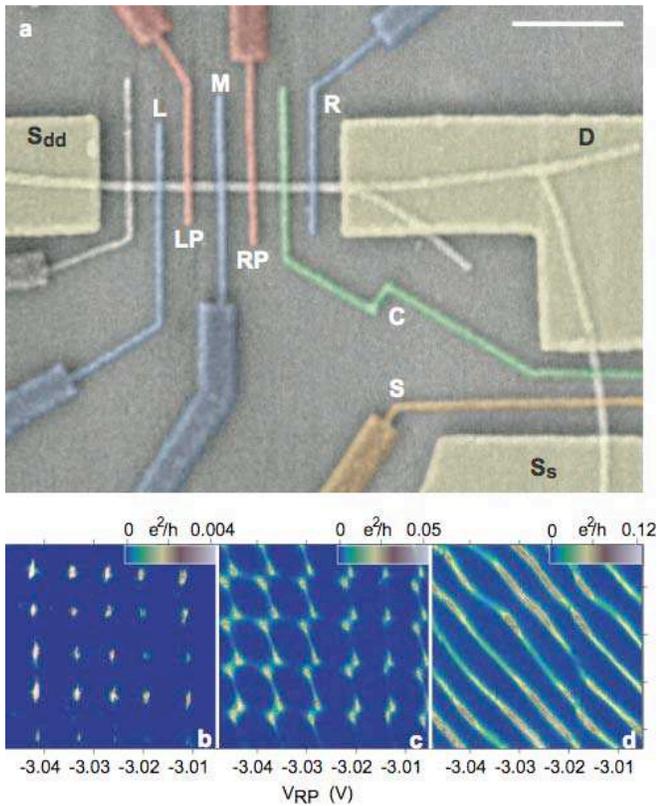}
\caption{\footnotesize{{\bf Ge/Si nanowire double dot device and demonstration of tunable interdot coupling.  a}  SEM image of the actual device used for all measurements.  
The double dot is formed with gates L, M, and R shown in blue, and the plunger gates LP and RP
(red) tune the energy levels of each dot. On an adjacent
nanowire, the charge sensor is a contact-defined single dot
capacitively coupled to the double dot with the coupler C (green).
The sensor is biased to the side of a Coulomb blockade peak using
gate S (orange).
The gate shown in gray was not used.
$S_{\rm dd}$, $S_{\rm s}$, and and $D$ label double dot source, sensor source and shared drain contacts, respectively.
Scale bar, 500 nm.
{\bf b}-{\bf d}  Differential conductance (color scale) is measured as a function of plunger voltages $V_{\rm LP}$ and $V_{\rm RP}$. 
With the side barriers fixed at $V_{\rm L}= -0.55$ V and  $V_{\rm R} = 0$
V,  changing the middle barrier voltage $V_{\rm M}$ shows three regimes of interdot coupling. {\bf b}
For weak interdot coupling ($V_{\rm M}= -0.72$ V), transport is allowed on
an array of triple points corresponding to resonant alignment of energy levels in the two dots with the chemical potential of the leads.
{\bf c} At intermediate coupling ($V_{\rm M} = -0.85$ V), cross-capacitance and tunneling
between dots split the triple points to create the honeycomb
charging pattern. {\bf d}  For strong coupling ($V_{\rm M}= -2.15$ V), an effective
single dot is formed, producing diagonal Coulomb blockade peaks.}}
\end{figure}
%-------------------%

\par
The potential of solid state spin qubits is underscored by the recent demonstration of coherent spin control in gate-defined double quantum dots (DQDs) with integrated charge sensors in GaAs two-dimensional electron gases (2DEGs) \cite{Petta-Sci05, Koppens-Nat06}.  
Additionally, few-electron InAs nanowire single and DQD devices allow access to strong spin-orbit interactions, which may prove useful for spin control \cite{Fasth-NL05,Pfund-condmat07, Fasth-condmat07}.  
In III-V materials, however, hyperfine coupling limits electron spin coherence. 
As a result, the prospect of long coherence times in group-IV materials due to the predominance of spin-zero nuclei \cite{Tyryshkin-PE06} has stimulated several proposals \cite{Kane-Nat98,Vrijen-PRA00,Friesen-PRB03,Trauzettel-NatP07} and significant experimental effort. 
Experimental progress in this direction includes realizations of DQDs in carbon nanotubes \cite{Biercuk-NL05,Sapmaz-NL06} and Si:P \cite{Chan-JAP06}, as well as single dots in Si and Ge/Si nanowires \cite{Zhong-NL05, Lu-PNAS05} and Si/Ge 2DEGs \cite{Klein-APL07,Berer-APL06,Sakr-APL05}.

\par
The chemically synthesized Ge/Si core/shell heterostructure nanowires (NWs) used here provide a high mobility one-dimensional hole gas with a mean free path on the order of hundreds of nanometers even at room temperature \cite{Lu-PNAS05}. 
The valence band offset of $\sim500$ meV between Ge and Si leads to a natural accumulation of holes in the Ge core, thus avoiding intentional impurity doping. 
These characteristics, along with the high controllability and reliability in material properties and device fabrication, make Ge/Si NWs ideal for coherent electronic devices.

%-----------------%
\begin{figure}
\center \label{figure2}
\includegraphics[width=3.4in]{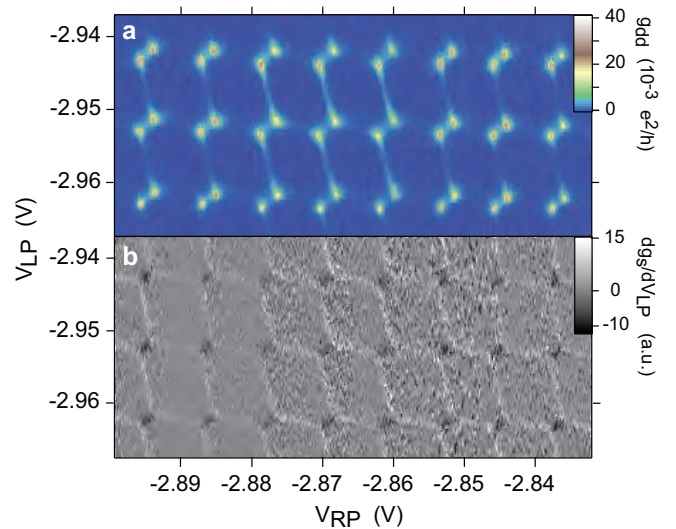}
\caption{\footnotesize{{\bf Simultaneous transport and charge sensing measurements.  a}  Double dot conductance ${\rm g_{dd}}$ as a function of gate voltages $\VLP$ and $\VRP$.  {\bf b} Simultaneously measured sensor dot conductance ${\rm g_s}$, differentiated with respect to gate voltage $\VLP$.}}
\end{figure}
%------------------%

\par
The DQD is formed by depleting a Ge/Si NW hole gas using metal gate electrodes. 
Three top gates, denoted L, M, and R in Fig.~1a, create barriers to define the dots, with the coupling between dots controlled by the middle barrier. 
Plunger gates LP and RP tune the energy levels of each dot. 
The device was measured in a dilution refrigerator with a base electron temperature of 150 mK using lock-in amplifiers and a current preamplifier (see Methods).

\par
Figures 1b--d show the differential conductance of the DQD, $g_{\rm dd}$, as a function of plunger voltages $\VLP$ and $\VRP$. 
With the side barrier voltages fixed at $\VL = -0.55$ V and $\VR = 0$ V, changing the middle barrier voltage $\VM$ shows three regimes of interdot coupling. 
For weak coupling ($\VM = -0.72$ V), transport occurs only at triple points where the energy levels of the two dots align with the chemical potential of the leads, resulting in a rectangular array of high conductance points (white spots in Fig.~1b). 
Setting $\VM$ to $-0.85$ V increases the coupling so that cross capacitance and tunneling between dots split the triple points to create the honeycomb charging pattern characteristic of DQDs \cite{Wiel-RMP03} shown in Fig.~1c. 
For strong coupling ($\VM = -2.15$ V), a single dot is effectively formed, producing the diagonal Coulomb blockade peaks seen in Fig.~1d. 
These data demonstrate fully tunable interdot coupling of the Ge/Si NW DQD. 
Additionally, extension of these techiniques should allow multiple quantum dots to be arbitrarily positioned along a NW, with independent electrical control over tunnel barriers and dot charges.

\par
Measuring the differential conductance of each single dot as a function of source-drain bias yields Coulomb diamonds (see Supplementary Information, Fig.~S1), from which we extract charging energies $E_C = e^2/C_{\Sigma}$ of 3.1 (2.6) meV for the left (right) dot, giving total capacitances $C_{\Sigma} \sim 52\  (61)$ aF. 
Single-particle level spacing $\sim250$ $\mu$eV was also measured from Coulomb diamonds. 
We place a lower bound of several hundred holes in each dot based on the number of charge transitions before tunnel rates inhibited further measurement.

%---------------------%
\begin{figure}[b]
\center \label{figure3}
\includegraphics[width=3.4in]{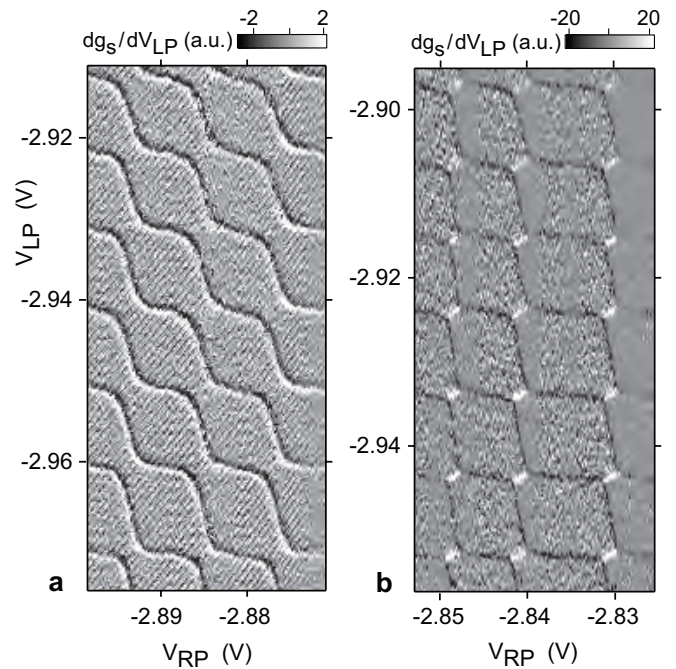}
\caption{\footnotesize{\textbf{Charge sensing of an isolated double dot.} 
Differentiated sensor conductance ${\rm dg_s/d\VLP}$ measured with the double dot weakly coupled to the leads (${\rm g_{dd}} < 10^{-5}\ e^2/h$) for {\bf a} strong ($\VM=-859$ mV) and {\bf b} weak interdot coupling ($\VM=-845$ mV).}}
\end{figure}
%----------------------%

\par
Key to realizing few-electron DQD devices in 2DEGs is the ability to noninvasively read out the charge state of the DQD, even when the tunnel coupling to the leads is too small to measure a current \cite{Elzerman-PRB03}. 
In one-dimensional systems, charge sensing was demonstrated recently in a carbon nanotube single dot using a radio-frequency single electron transistor \cite{Biercuk-PRB06}. 
Here, we have developed a novel approach to charge sensing by capacitively coupling a single dot on an adjacent nanowire to the DQD. 
This method provides a simpler alternative in terms of fabrication and in addition may offer a means of easily integrating elements in a larger-scale circuit.  
As shown in Fig.~1a, the charge sensor is a contact-defined single dot capacitively coupled to the DQD with the coupler C (green).  
The sensor plunger gate S is used to bias the sensor dot to the side of a Coulomb blockade peak for maximum sensitivity to changes in the number of holes on the DQD (see Supplementary Information, Fig.~S2).
The charge sensor operates by gating the sensor dot with the changes in electrostatic potential associated with charge transitions in the DQD. 

\par
To test the capabilities of the charge sensor, we have made simultaneous transport and charge sensing measurements in the intermediate coupling regime ($\VM = -0.86$ V). 
Figure 2a shows $g_{\rm dd}$ as a function of gate voltages $\VLP$ and $\VRP$, displaying a honeycomb pattern similar to Fig.~1c. 
We note that the well-developed honeycomb pattern is a consequence of the extremely clean and highly tunable Ge/Si hole gas system. 
Figure 2b shows the sensor conductance, $g_{\rm s}$, which was measured simultaneously with $g_{\rm dd}$ and numerically differentiated with respect to $\VLP$. 
With the sensor dot biased on the negative slope of a Coulomb blockade peak, transfer of a hole from one dot to the leads results in a step up in conductance due to the decreased electrostatic potential in the sensor dot. 
On the other hand, transfer of a hole from the left dot to the right dot results in a step down in conductance because the sensor coupler is closer to the right dot. 
These steps up and steps down are observed in Fig.~2b as bright and dark features, respectively. 
The bright features in Fig.~2b highlight the ability of the charge sensor to detect transitions that are only weakly present in the transport signal as cotunneling lines. 

\par
To demonstrate the advantage of the charge sensor to probe regimes inaccessible to transport measurements, we next decouple the DQD from the leads by setting $\VL = 0$ mV and $\VR = 250$ mV so that $g_{\rm dd} < 10^{-5} e^2/h$. 
Figures 3a, b show $dg_s/d\VLP$ for both (a) strong and (b) weak interdot coupling. 
Clear honeycomb charging patterns are seen in both cases. 
In Fig.~3a the sensor dot is biased near the top of a Coulomb blockade peak such that $g_s$ responds nonlinearly both to the charge transitions on the DQD and to the compensation (see Methods) applied to gate S, resulting in a peak in $g_s$ rather than a step. 
Thus $dg_s/d\VLP$ is both positive and negative across a transition. 
We also note that the sign of $dg_s/d\VLP$ in Fig.~3b is reversed relative to that in Fig.~2b because the sensor dot is biased to the opposite side of a Coulomb blockade peak.

%----------------------%
\begin{figure}
\center \label{figure4}
\includegraphics[width=3.4in]{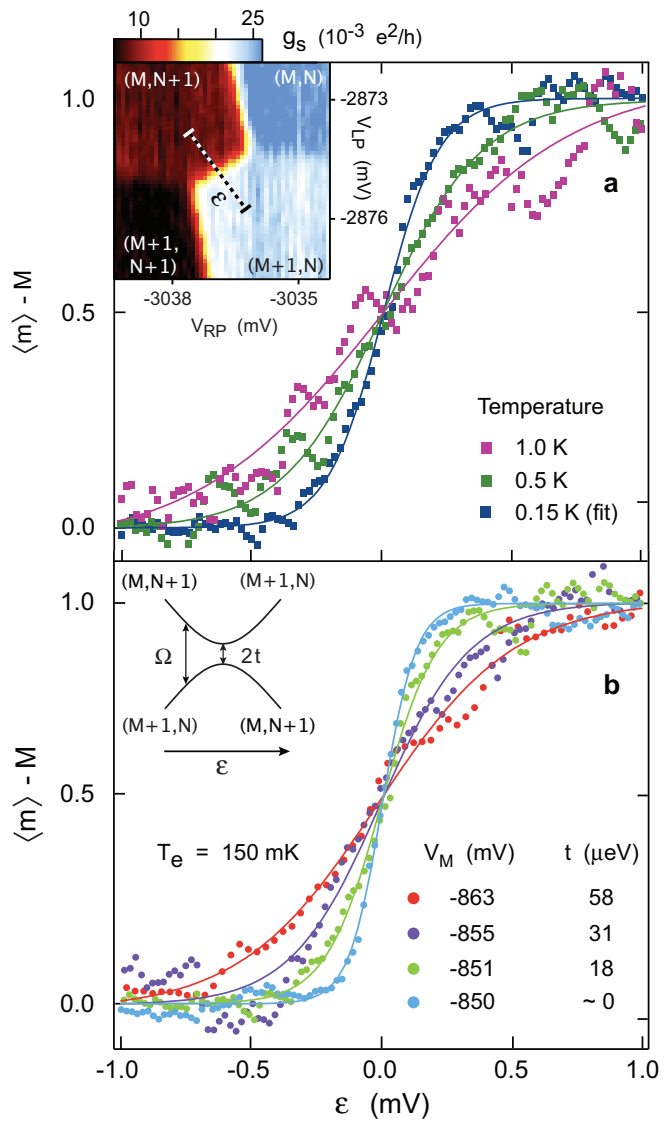}
\caption{\footnotesize{{\bf Interdot tunneling measured with charge sensor.  a} Sensor conductance ${\rm g_s}$ rescaled to reflect excess charge (in units of $e$) on the left dot along the detuning diagonal $\epsilon$ (dotted line in inset shows $\epsilon = -1\ {\rm to}\ 1$ mV) at $T_e$ = 0.15 K (dark blue), 0.5 K (dark green), and 1.0 K (pink) for $V_{\rm M} = -851$ mV.
The solid lines are fits to equation (\ref{twolevel}).
Inset:  sensor conductance ${\rm g_s}$ showing the charge stability diagram in the region used for {\bf a} and {\bf b}.
The charge state with M (N) holes on the left (right) dot is denoted (M,N).
Average values of ${\rm g_s}$ are 6.5, 8.4, 23, and 26 $\times\ 10^{-3}\ e^2/h$ on the black, red, white, and blue plateaus, respectively.
{\bf b}  Excess charge on the left dot (${\rm g_s}$, rescaled) at base temperature for several values of $V_{\rm M}$.
The temperature-broadened curve (blue) widens as $V_{\rm M}$ is made more negative, increasing the tunnel coupling $t$ which is extracted from fits to Eq.~(\ref{twolevel}) (solid lines).
The fit to the temperature-broadened curve gives a base electron temperature of 150 mK, in agreement with Coulomb blockade peak widths.
Inset:  schematic energy diagram of the two-level system model, showing the splitting between ground and excited states as a function of detuning $\epsilon$ with an anticrossing of $2t$ at $\epsilon=0$.}}
\end{figure}
%-----------------------%

\par
Significantly, charge sensing also allows access to interdot transitions at fixed total charge which are difficult to study in transport measurements \cite{DiCarlo-PRL04}.
Following the ``detuning" diagonal $\epsilon$ between a triple point pair (dotted line in inset to Fig.~4) from negative to positive transfers charge from the right dot to the left dot, resulting in a conductance step on the sensor dot.
Denoting by $(M,N)$ the charge state with $M$ ($N$) holes on the left (right) dot, we model the sensor conductance across the transition from $(M+1,N)$ to $(M,N+1)$ in terms of an isolated two-level system in thermal equilibrium~\cite{DiCarlo-PRL04}.
When the tunnel coupling $t$ mixing the two states is small relative to the single-particle level spacings for the individual dots, sensor conductance depends on detuning $\epsilon$ as
\begin{equation}\label{twolevel}
{\rm g_s} = {\rm g_0} + \delta{\rm g}\, \frac{\epsilon}{\Omega}\,\tanh\!\left(\frac{\Omega}{2k_BT_e}\right),
\end{equation}
where $\Omega=\sqrt{\epsilon^2+4t^2}$ is the energy splitting between the ground and excited states.
Rescaling the sensor conductance so that ${\rm g_0}=\delta {\rm g} = 1/2$ yields the excess charge on the left dot, $\langle m\rangle - M$.
Measurements of excess charge versus detuning are plotted in Figs.~4a, b.
The model given by equation (1) is in good agreement with the data as shown by the solid lines in Figs.~4a, b.

\par
At low temperatures the electrons in the device can be at a higher temperature than our thermometry.
Because the transition width depends on both temperature and tunneling, we first calibrate the electron temperature by measuring the transition at elevated temperatures where the electrons are well thermalized, as shown in Fig.~4a.
Data at the highest temperatures (0.75 and 1.0 K) allow extraction of the lever arm which is used to estimate a base electron temperature of 150 mK for the blue curve in Fig.~4a, which is also in agreement with Coulomb blockade peak widths.
With the temperature calibrated, we now examine the sensing transition as a function of interdot tunneling in the regime of strong tunneling, $t\gtrsim k_BT_e$.
Figure 4b shows excess charge along the detuning diagonal for several values of $\VM$ at base temperature.
For $\VM=-850$ mV the transition did not narrow for less negative $\VM$ so we assume the transition is thermally broadened with $t\sim0$.
For the more negative values of $V_M$, fixing $T_e = 150$ mK allows extraction of the tunnel couplings $t$ as the only free parameter in fits to equation (\ref{twolevel}).
The noise in ${\rm g_s}$ is dominated by fluctuations in the bias point up and down the Coulomb blockade peak and is significantly reduced by averaging.
Each of the curves in Figs.~4a, b is an average of 100 sweeps, and the inset to Fig.~4a is an average of 35 two-dimensional scans.

\par
In conclusion, we have demonstrated a fully tunable DQD in a Ge/Si heterostructure NW using local gate electrodes. 
We also presented a novel approach to charge sensing by capacitively coupling the DQD to a single dot on an adjacent NW.  
The integration of these two components provides the capabilities required for future devices to access the few-charge regime and carry out coherent spin manipulation experiments.  
The prospects of forming spin qubits with Ge/Si NW DQDs are bright. 
Long spin coherence times are expected to result from suppressed hyperfine interactions due to the absence of nuclear spin.  
Because of strong spin-orbit interactions in the valence band, hole spin lifetimes are in general much shorter than those of electrons \cite{Bulaev-PRL05}, but in our system quantum confinement and strain-induced splitting of the heavy-hole and light-hole subbands may reduce spin-orbit interactions \cite{Tyryshkin-PE06,Schaffler-SST97}. 
Furthermore, the observed ambipolar behavior in these NWs \cite{Lu-PNAS05} ensures electron and hole conduction could be easily tuned electrostatically and suggests the possibility of studying electron and hole spins in the same device. 
This clean, highly controllable system offers a promising route to studies of coherent electronic devices free of nuclear spin. 
%-----------------------%
\section{Methods}
\subsection{Fabrication of Ge/Si NW Devices}
The undoped Ge/Si core/shell NWs were grown via a two-step chemical
vapor deposition process \cite{Lu-PNAS05}. 
The nanowires have an average core diameter of 14.6 nm and Si shell thickness of 1.7 nm,
and normally exhibit $\langle110\rangle$ growth direction. 
AFM measurements of the nanowires forming the actual device presented here indicate $\sim\!15$ nm diameter for the DQD NW and $\sim\!10$ nm diameter for the sensor NW.
The degenerately doped Si substrate with 600 $\mu$m thermal oxide served as a global backgate and was set to $-2$ V for all measurements.
All source-drain contact electrodes (50 nm Ni) were defined by electron-beam
lithography and deposited by thermal evaporation. 
Transparent contacts were obtained for the DQD NW, while contact barriers for the sensor nanowire formed a dot at low temperature, possibly due to its smaller diameter or to a thicker native oxide layer on the shell. 
The NWs and source-drain electrodes were then covered with a 12 nm Hf${\rm O_2}$
high dielectric constant layer ($\kappa\sim23$) using atomic layer deposition. 
Hf${\rm O_2}$ was deposited at 110 $^{\circ}$C in 100 cycles of 1 s water vapor pulse, 5 s ${\rm N_2}$ purge, 3 s precursor, and 5 s ${\rm N_2}$ purge.
Tetrakis (dimethylamino) hafnium [Hf(N(CH$_3$)$_2$)$_4$] was used as precursor.
Electron-beam lithography was used to define the top gates, followed by thermal evaporation of Al (50 nm).
Top gates were approximately 30 nm wide with 110 nm spacing.

\subsection{Measurements}
An ac excitation of 10 $\mu$V was applied to the source contacts of the DQD and sensor at 149 and 109 Hz, respectively.
The shared drain contact was connected to a current preamplifier, followed by separate lock-in amplifiers to measure the DQD conductance $g_{\rm dd}$ and the sensor conductance $g_{\rm s}$. 
To cancel the cross-coupling between gates and maintain the sensor in a high-sensitivity position, the sensor plunger voltage $V_{\rm S}$ was adjusted during sweeps of $\VLP$ and $\VRP$.

%----------------------%
\subsection{Acknowledgments}
We thank L.~DiCarlo and E.~A.~Laird for experimental assistance and helpful discussions.  CML acknowledges support from DARPA and Samsung Electronics. CMM acknowledges support from DTO. HOHC acknowledges support from the NSF.

\end{document}